\documentclass[12pt]{elsarticle}
\usepackage[latin1]{inputenc}
\usepackage{lineno}
\usepackage{graphicx}
\usepackage{amsmath}
\usepackage{amssymb}

\newcommand{\tm}[1]{\raisebox{1ex}{\scriptsize TM}}

\begin{document}

\begin{frontmatter}


\title{Realtime calibration of the A4 electromagnetic lead fluoride (PbF$_2$) calorimeter}

\author{S.~Baunack$^{\,a,}$\corref{cor1}}\ead{baunack@kph.uni-mainz.de}
\author{D.~Balaguer~R{\'i}os$^{\,a}$}
\author{L.~Capozza$^{\,a}$}
\author{J.~Diefenbach$^{\,a,}$}
\author{R.~Frascaria$^{\,b}$}
\author{B.~Gläser$^{\,a}$}
\author{D.~v.~Harrach$^{\,a}$}
\author{Y.~Imai$^{\,a}$}
\author{R.~Kothe$^{\,a}$}
\author{R.~Kunne$^{\,b}$}
\author{J.~H.~Lee$^{\,a,***}$}
\author{F.~E.~Maas$^{\,a}$}
\author{M.~C.~Mora Esp\'{i}$^{\,a}$}
\author{M.~Morlet$^{\,b}$}
\author{S.~Ong$^{\,b}$}
\author{E.~Schilling$^{\,a}$}
\author{J.~van~de~Wiele$^{\,b}$}
\author{C.~Weinrich$^{\,a}$}
\address{$^a\,$Institut für Kernphysik, Johannes-Gutenberg-Universität, 55099 Mainz, Germany}
\address{$^b\,$Institut de Physique Nucl\'{e}aire, CNRS-IN2P3, Universit\'{e} Paris-Sud, F-91406 Orsay Cedex, France}
\cortext[cor1]{Corresponding author. Address: Institut f{\"u}r Kernphysik,
  J.J. Becherweg 45, 55099 Mainz, Germay. Fax: +4961313922964, Phone: +4961313925808}

\begin{abstract}

Sufficient energy resolution is the key issue for the calorimetry in particle
and nuclear physics. The calorimeter of the A4 parity violation experiment at
MAMI is a segmented calorimeter where the energy of an event is determined by
summing the signals of neighbouring channels. In this case the precise
matching of the individual modules is crucial to obtain a good energy
resolution. We have developped a calibration procedure for our total absorbing
electromagnetic calorimeter which consists of 1022 lead fluoride (PbF$_2$)
crystals. This procedure reconstructs the the single-module contributions to
the events by solving a linear system of equations, involving the inversion of
a 1022$\times$1022-matrix. The system has shown its functionality at beam
energies between 300 and 1500 MeV and represents a new and fast method to keep
the calorimeter permanently in a well-calibrated state.
\end{abstract}

\begin{keyword}
calorimeters \sep calibration

\PACS 
07.20.Fw \sep 
06.20.fb 

\end{keyword}

\end{frontmatter}


\newpage
\section{Introduction}
The calibration procedure described here has been developed for the total
absorbing homogenous electromagnetic calorimeter of the A4 experiment at the
Mainz electron accelerator facility MAMI. The A4 collaboration investigates
single spin asymmetries in the cross section of the elastic scattering of
polarized electrons off unpolarized hydrogen and
deuterium~\cite{Maas04a,Maas04b,Maas04c,Baunack09}. In order to separate
elastic scattering events from inelastic background, the energy of the
detected particles is measured by means of a totally absorbing electromagnetic
calorimeter. For such a a measurement, the analog signals of nine neighbouring
modules are summed to form the event signal which is then digitized. It is
obvious that a coherent signal response of all calorimeter modules is decisive
to obtain an optimal energy resolution.

For the calibration of an electromagnetic calorimeter various methods are
applicable. One common method is to use a radioactive source which delivers
well-known energies up to a few MeV~\cite{CMS08}. For a total absorbing
calorimeter which measures energies in the range between several hundred MeV
and a few GeV, however, no suitable radioactive sources exist. Another
possible procedure uses direct photoelectrons. Light produced by UV lasers or
LEDs is guided to the calorimeter modules by fibers~\cite{D098}. However, the
precision of this method is limited by the uniformity of the coupling of the
fibers to the calorimeter crystals, the homogenity of the light sources and the
losses in the individual fibers that determines the light input variation from crystal
to crystal. A third common technique is the use of cosmic events for calibration
purposes~\cite{Kloe09}. 

The method presented here is based on a calibration using elastically
scattered electrons which are easy to identify and have well-defined
energies. A difficulty arises from the fact that the data used to determine
the current calibration state are not individual signals of single
modules. Only the sum of the signals of nine neighbouring modules is
accessible. The reconstruction of the individual yields of the modules
involves linear algebra and is one of the main tasks in the calibration
procedure.

\section{The electromagnetic homogenous lead fluoride calorimeter}
\begin{figure}
\begin{center}
\includegraphics[scale=.7]{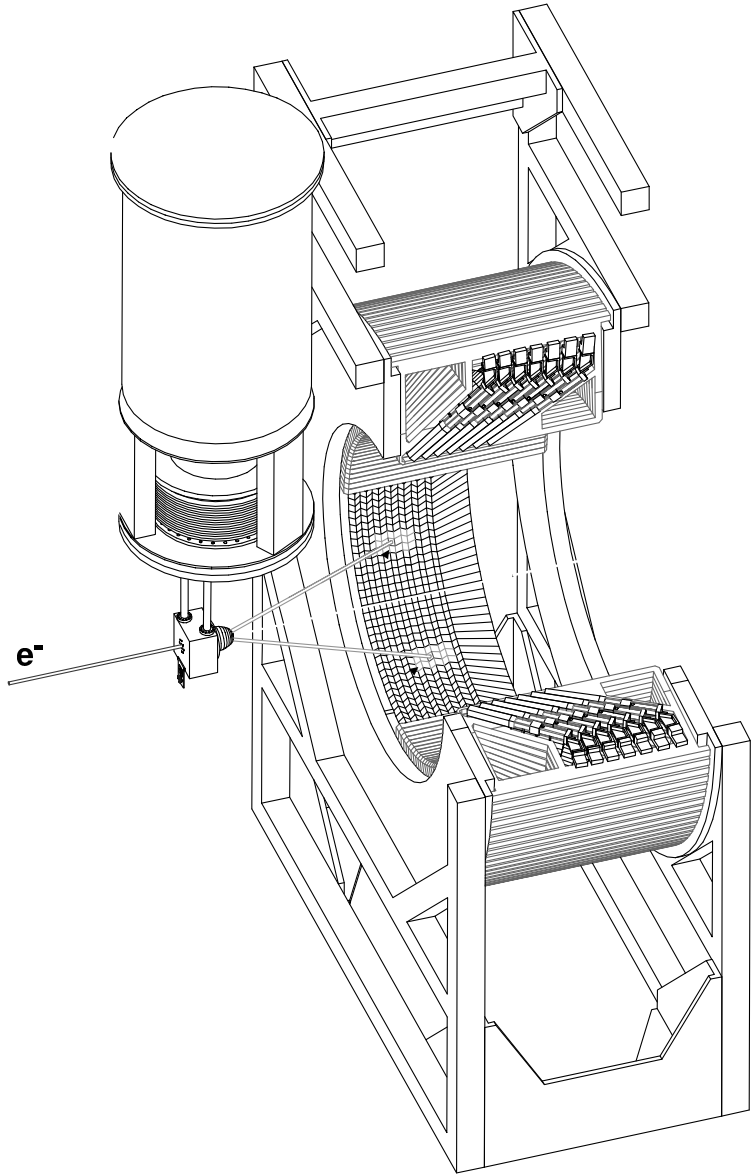}\caption{\label{fig.calorimeter}Partly
  cut-away view of the lead fluoride calorimeter. The electrons come from the
  left, hit the hydrogen target and are scattered into the calorimeter at
  polar angles of $30^\circ\leq\theta_e\leq 40^\circ$. The resulting
  electromagnetic shower is contained within a cluster of 3$\times$3
  PbF$_2$ crystals.}
\end{center}
 \end{figure}
\begin{figure}
\begin{center}
\includegraphics[scale=1]{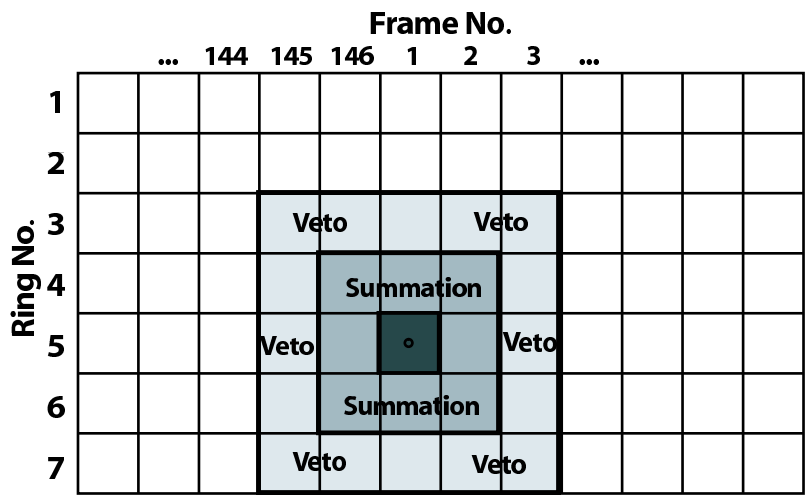}\caption{\label{fig.Vetozone}Schematic
  view of the array of lead fluoride crystals in the calorimeter. The modules
  are labelled according to the aluminium frame they are mounted on (1 to 146)
  and the ring they belong to (1 to 7). In this example, a particle hits the
  calorimeter at frame 1, ring 5 (marked in dark gray). In order to determine
  the energy of this event, the signal of this module and the signals of the
  eight neighbouring modules are summed up (medium gray). During the
  integration time no other particle may hit the area around this cluster
  (light gray), because the electromagnetic shower of this event would overlap
  with that from the first particle. In the case of such a pile-up both events
  are rejected.}
\end{center}
\end{figure}
\begin{figure}
\begin{center}
\includegraphics[scale=0.82]{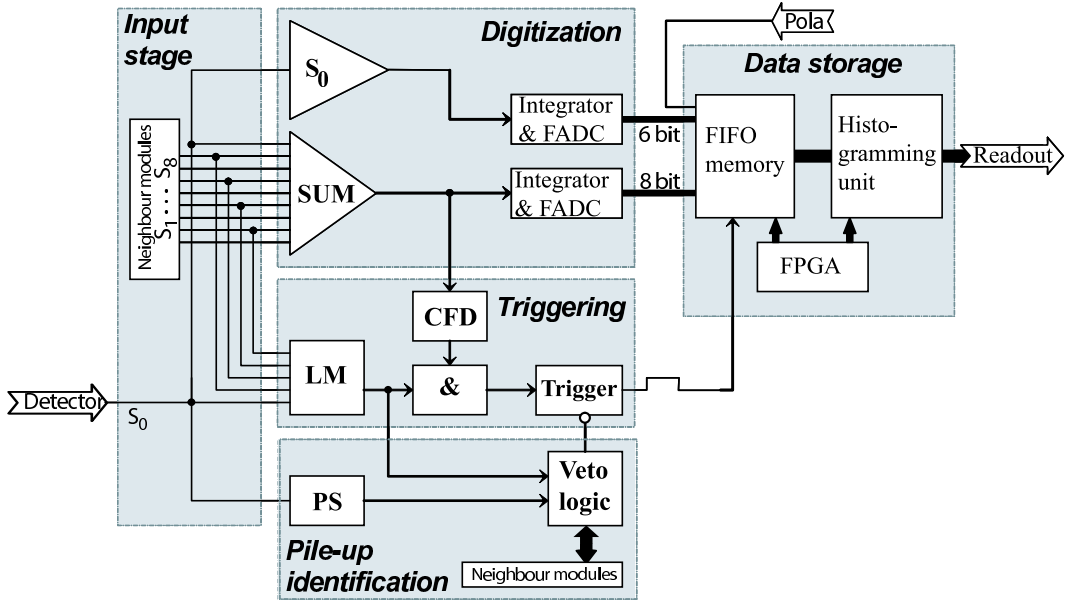}\caption{\label{fig.electronics}Diagram
  of the readout electronics~\cite{Kothe08}. The input signal from the
  photomultiplier tube is denoted with $S_0$. The signals of the neighbouring
  channels are denoted with $S_1$ to $S_8$. The signals are summed up by a
  summation amplifier. A triggering and a veto unit decide whether the event is
  stored into the histograms.}
\end{center}
\end{figure}

The calorimeter consists of 1022 individual lead fluoride (PbF$_2$) crystals
mounted on 146 aluminium frames and arranged in 7~rings
(fig.~\ref{fig.calorimeter}). PbF$_2$ is a pure Cherenkov radiator with high
transmission down to 270~nm~\cite{Achenbach01}. Its radiation length is
$X_0=0.93$~cm and its effective ``Moli{\`e}re radius'' for the production of
Cherenkov photons from an electromagnetic shower is $R_M=1.8$~cm. The light is
read out by photomultiplier tubes XP~2900 with borosilicate windows and
transistorized, actively stabilized bases.  We employ seven types of
trapezoid-shaped PbF$_2$ crystals with slightly different shapes. The
calorimeter geometry is that of a barrel with the individual crystals pointing
to the target. The crystals have a cross section of 2.6~cm x 2.6~cm at the
front, 3.0~cm x 3.0~cm at the back, and a length of 15--18~cm. In longitudinal
direction we have chosen the crystal length to cover more than 15 radiation
lengths in order to minimize shower leakages. In the transverse direction we
have chosen a crystal width of 4/3 $R_M$. Hence, more than 95\% of the
Cherenkov light of an event is emitted in a cluster of 3$\times$3
crystals. The signals of all channels within this cluster are summed up for
the energy measurement of an event (fig.~\ref{fig.Vetozone}). If another
particle hits the same crystal (dark gray) during the integration window of
20~ns, a veto logic detects this double hit as long as the time lag between
the two events is at least 6~ns. Also, during the integration time no other
particle may hit the area around this cluster (light gray), because the
electromagnetic shower of this event would overlap with that from the first
particle. Both types of pile-up events would lead to a wrong energy
determination and are rejected.
Triggering, pile-up rejection, summing up the signals and storing them at
event rates of up to 100~MHz requires parallelized electronics~\cite{Koebis98}. For each of the
1022~channels of the calorimeter there is a dedicated electronic circuit
(fig.~\ref{fig.electronics}) implementing the event detection as follows: when
a particle hits the calorimeter and deposits energy in the crystals around the
impact zone, the trigger electronics of each of those modules recognizes the
hit in the calorimeter. The center of the impact is determined by finding the
crystal with the largest signal pulse height ({\sl Local Maximum, LM}). This
crystal defines the center of a 3$\times$3 cluster of crystals. The signals of the
photomultiplier tubes of these nine crystals are summed up by an analog
summation circuit. If this sum exceeds the trigger level of a constant fraction
discriminator ({\sl Constant Fraction, CFD}) and there is neither a double hit
in the central crystal recognized by the pulse shaper unit ({\sl Pulse Shaper,
  PS}) nor a second hit in the neighbouring modules, then a trigger flag is
set. The signal of the 3$\times$3-cluster is digitized by an 8-bit ADC, the signal of
the central crystal is additionally digitized by a 6-bit ADC and both values are
stored in a histogramming unit. After a specified run time --- usually 5
minutes --- the histogram is read out and transferred to a storage device.

\section{The calibration procedure}
\begin{figure}
\begin{center}
\includegraphics[scale=1]{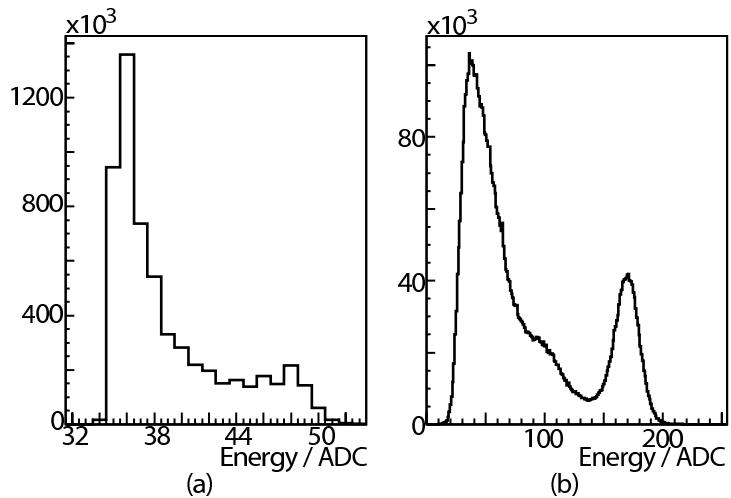}\caption{\label{fig.Spectra}Typical
  energy spectra of the PbF$_2$ calorimeter: (a) Histogram of the 6-bit ADC
  with the signal of a single module, (b) Histogram of the 8-bit ADC with the
  signals of the modules of a 3$\times$3-cluster summed up. One can identify
  the peak of the elastically scattered electrons which corresponds to an
  energy of 734~MeV at around ADC channel 48 in (a) and around ADC channel 170
  in (b). The cutoff at low ADC values is a threshold effect in the
  electronics. Due to the limited resolution of the 6-bit histogram we use the
  8-bit histograms for the analysis where the signals of a 3$\times$3-cluster
  were summed up.}
\end{center}
\end{figure}

The aim of the calibration is to normalize the gain for all 1022~modules of
the calorimeter: the same amount of energy deposited in a PbF$_2$ crystal
should result in the same integrated current, i.e. collected charge, in the
photomultiplier tubes (PMT) for all modules. Prior to installing the PMTs into
the calorimeter, all of them were precalibrated in a test stand. The transit
times were measured and the signal cable lengths adjusted to ensure an equal
timing. Also, precision electronic parts were used in the summation circuits,
e.g. resistors with 0.1\% tolerance and capacitors with 1\% tolerance. Even
with low tolerance parts fitted into the calorimeter, the individual gains may
still differ because of different light yields of the individual lead fluoride
crystals or because of differences in the optical couplings between crystal
and PMT. These residual differences are eliminated by adjusting the gains of
the individual modules by varying the high voltage of the PMT. In order to
determine these gains, we use the energy spectra which are measured during
regular data taking. During data taking, there are two types of histograms available: The
histograms from the 6-bit ADC which give the energy information of the single
modules and the histograms from the 8-bit ADC which contain the summed energy
information of a 3$\times$3-cluster of crystals. The histograms shown in
fig.~\ref{fig.Spectra} are examples of such spectra. At large ADC values one
can see the peak of elastic scattering events.
The cutoff at low ADC values is a threshold effect in the
electronics. The offsets are measured in a separate procedure~\cite{Kothe08}
and are located around ADC channel 32 for the 6-bit histogram and between ADC
channels -25 and +10 for the 8-bit histograms. Due to the limited resolution
of the 6-bit histogram we use the 8-bit histograms for the analysis. Since the
energy of the electrons after elastic scattering is fixed by the kinematics,
the position of the elastic peak given in ADC units together with the measured
offset allows an unambiguous determination of the gain of a 3$\times$3-cluster
in units of ADC/MeV. In order to disentangle the contributions of the single
modules to this sum signal, one needs to understand how the energy of an event
is deposited within the cluster. With the knowledge of the average lateral
distribution of the electromagnetic shower, one can set up a system of
equations which relates the sum signals to the individual gains of the single
modules. By solving this system of equations, one can determine the individual
amplification factors. The latter can be used to calculate a new set of high
voltages for the photomultipliers necessary to reach the desired calibration
state.

The following steps are necessary for the calibration
procedure. They will be discussed in detail in the remainder of this chapter:
\begin{enumerate}
\item Understand and parameterize the lateral distribution of the
  electromagnetic shower in the PbF$_2$ calorimeter.
\item Analyze the 1022 energy spectra to find the position of the elastic
  peak which gives the information about the signal gain for a whole 3$\times$3
  cluster
\item Using the methods of linear algebra, solve the equations and calculate the amplification factors of the
  individual single modules   
\item Calculate and apply new high voltages to reach the desired amplification
  for all modules
\end{enumerate}  

\subsection{Lateral distribution of the electromagnetic shower in the calorimeter}
\begin{figure}
\begin{center}
\includegraphics[scale=.65]{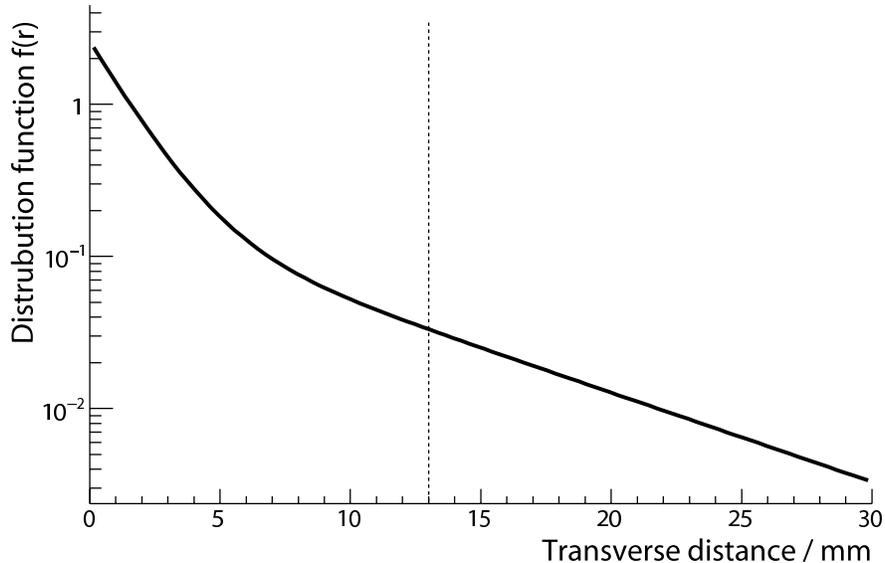}\caption{\label{fig.LateralShower}Lateral
  distribution (eq.~\ref{gl:modell}) of the energy of an electromagnetic
  shower in lead fluoride, where the parameters $a_1$, $a_2$, $b_1$ and $b_2$
  were determined by a GEANT simulation. The dotted line indicates the limit
  of the crystal surface whereas the origin of the x-axis lies on the center
  of the crystal.}
\end{center}
\end{figure}

When a scattered electron with an energy of several hundred MeV hits a
calorimeter module, it will lose its kinetic energy developing an
electromagnetic shower where bremsstrahlung and pair production are the
dominant processes. Since the crystals have a length of at least 15 radiation
lengths, electrons with energies up to 1~GeV will be absorped completely in
the calorimeter and there is no concern about the longitudinal distribution of
the shower. The transverse development of the electromagnetic shower scales
with the Moli\`ere radius $R_M$. About 90\% of the energy is deposited within
a cylinder with radius $R_M$. The lateral energy distribution of the
electromagnetic shower can be parameterized using a sum of two
exponentials~\cite{Bia89}:
\begin{linenomath}
\begin{equation}
f(r) = a_1 \exp\left(-r/b_1\right) + a_2 \exp\left(-r/b_2\right)
\label{gl:modell}
\end{equation}
\end{linenomath}
For our calorimeter material, lead fluoride, these coefficients have been
determined by GEANT simulations~\cite{Gri96}. Normalizing $f(r)$ to one,
\begin{linenomath}
\begin{equation}
\int\limits_{0}^{2\pi}\int\limits_0^{\infty}rf(r)\,d\phi\,dr=1
\end{equation}
\end{linenomath}
one gets $a_1=2.41~\mathrm{mm}^{-2}$, $a_2=0.19~\mathrm{mm}^{-2}$,
$b_1=1.5~\mathrm{mm}$ and $b_2=7.4~\mathrm{mm}$ (fig.~\ref{fig.LateralShower}).
With this shower distribution function one can calculate the amount of energy
that is deposited on average in the modules of a 3$\times$3 cluster whose central crystal has
been hit by a scattered particle. There are three ``types'' of crystals in
such a cluster~(fig.~\ref{fig.cluster}):
\begin{figure}
\begin{center}
\includegraphics[scale=.5]{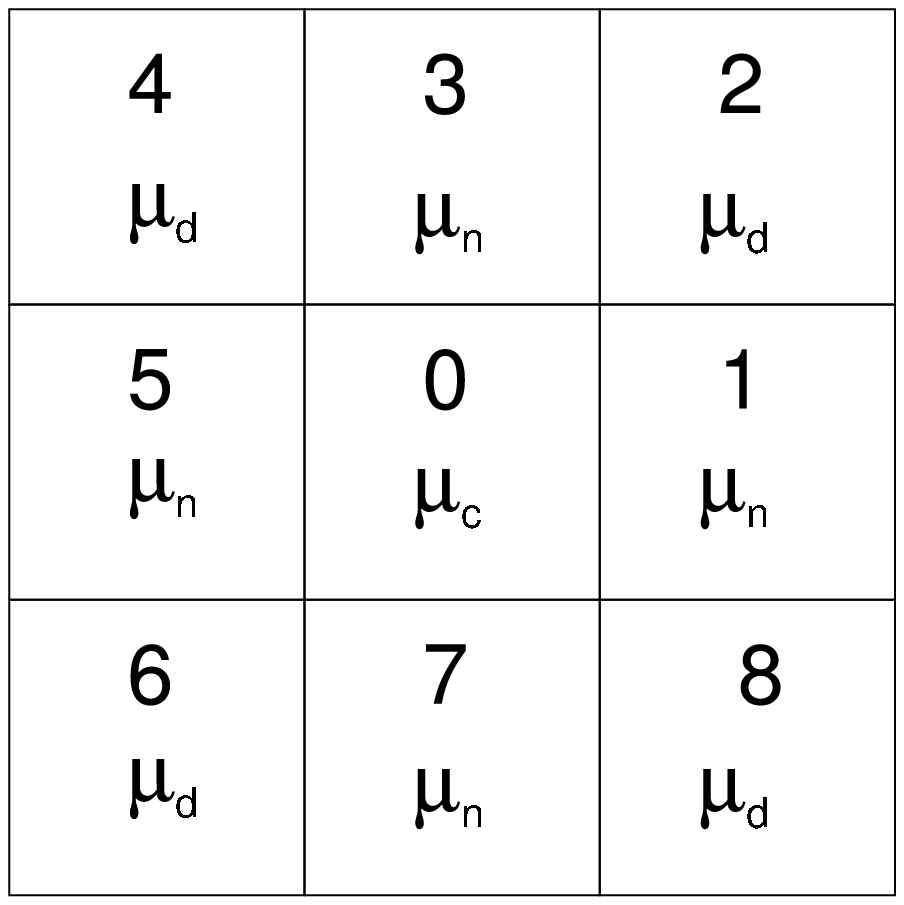}\caption{\label{fig.cluster}Schematic representation of
  a 3$\times$3 cluster of crystals. There are three ``types'' of modules, the
  central crystal (no. 0) which defines the 3$\times$3 cluster, the direct neighbours
  (no. 1,3,5,7) and the diagonal neighbours (no. 2,4,6,8). The fraction of
  energy that is deposited on average in these three classes of crystals is denoted by
  $\mu_c$, $\mu_n$ and $\mu_d$ respectively.}
\end{center}
\end{figure}
the central crystal {\sl c} (no. 0) which defines the 3$\times$3 cluster, the
direct neighbours {\sl n} (no. 1,3,5,7) and the diagonal neighbours {\sl d}
(no. 2,4,6,8). How the energy is spread out over the modules for a single
event depends on the impact position of the incident particle as well as on
the actual development of the electromagnetic shower which is subject to
statistical fluctuations. However, since the calibration is performed using a
histogram of a large number of events (between $10^6$ and $10^7$ events), only the
average of that distribution over all possible impact positions and over a
large number of electromagnetic showers is relevant. The fraction of energy
that is deposited on average in these three classes of crystals is denoted by
the distribution parameters $\mu_c$, $\mu_n$ and $\mu_d$, respectively:
\begin{linenomath}
\begin{equation}
  E_{\rm{deposit}}=\left(\mu_c+4\mu_n+4\mu_d\right)E_{\rm{incident}}
\end{equation}
\end{linenomath}
 These parameters can be determined by averaging the energy deposition of the
 electromagnetic shower over all possible impact positions on the central
 crystal's surface. We assume that within one crystal all impact positions
 have equal probability which is a good approximation considering the small
 single-crystal acceptance:
\begin{linenomath}
\begin{equation}
\mu_{c,n,d}=\frac{1}{S_c}\,\int\limits_{S_c}\;\:\int\limits_{S_{c,n,d}}f(x'-x,y'-y)\,dx'\,dy'\,dx\,dy 
\label{eq.DistributionPara}
\end{equation}
\end{linenomath}
The $S_{c,n,d}$ denote the surface of the center, neighbour or diagonal
crystal, respectively, $f(x,y)=f(r)$ with $r=\sqrt{x^2+y^2}$ parameterizes the lateral energy deposition of the
electromagnetic shower (eq.~\ref{gl:modell}). We have computed this integral
using the actual dimensions of the crystals. The parameters are $\mu_c=65.3\%$,
$\mu_n=6.2\%$ and $\mu_d=1.2$\%. On average, the biggest fraction of energy is
deposited in the central crystal, while the diagonal neighbours contribute
only 1.2\% each.\\
\begin{figure}
\begin{center}
\includegraphics[scale=1]{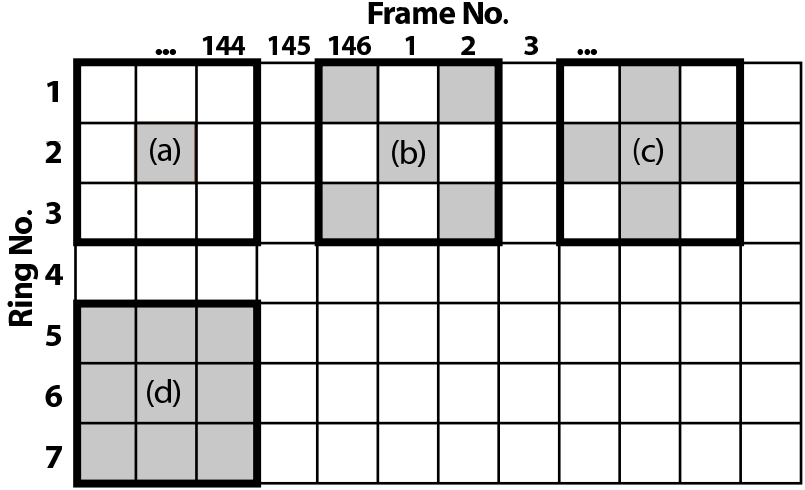}\caption{\label{fig.MueMessSchema}Method for a measurement of the
  distribution parameters $\mu$ with the electron beam. The colored boxes
  indicate modules which are supplied with high voltages: (a) Only the central
  crystal, (b) the central crystal and its
  diagonal neighbours, (c) the central crystal
  and its direct neighbours, (d) all modules
  of a 3$\times$3 cluster.}
\end{center}
\end{figure}
We also performed a measurement of the distribution parameters
$\mu_{c,n,d}$. The method is to supply only the PMTs of selected modules of a
3$\times$3-cluster with high voltages (fig.~\ref{fig.MueMessSchema}). In this
way the other modules, either the direct neighbours or the
diagonal neighbours or all neighbours, do not contribute to the signal,
resulting in a shift of the position of the elastic peak in the ADC spectra
(fig.~\ref{fig.muebestimm}). Measuring this shift, the contribution of
neighbouring modules can be determined.
\begin{figure}
\begin{center}
\includegraphics[scale=0.6]{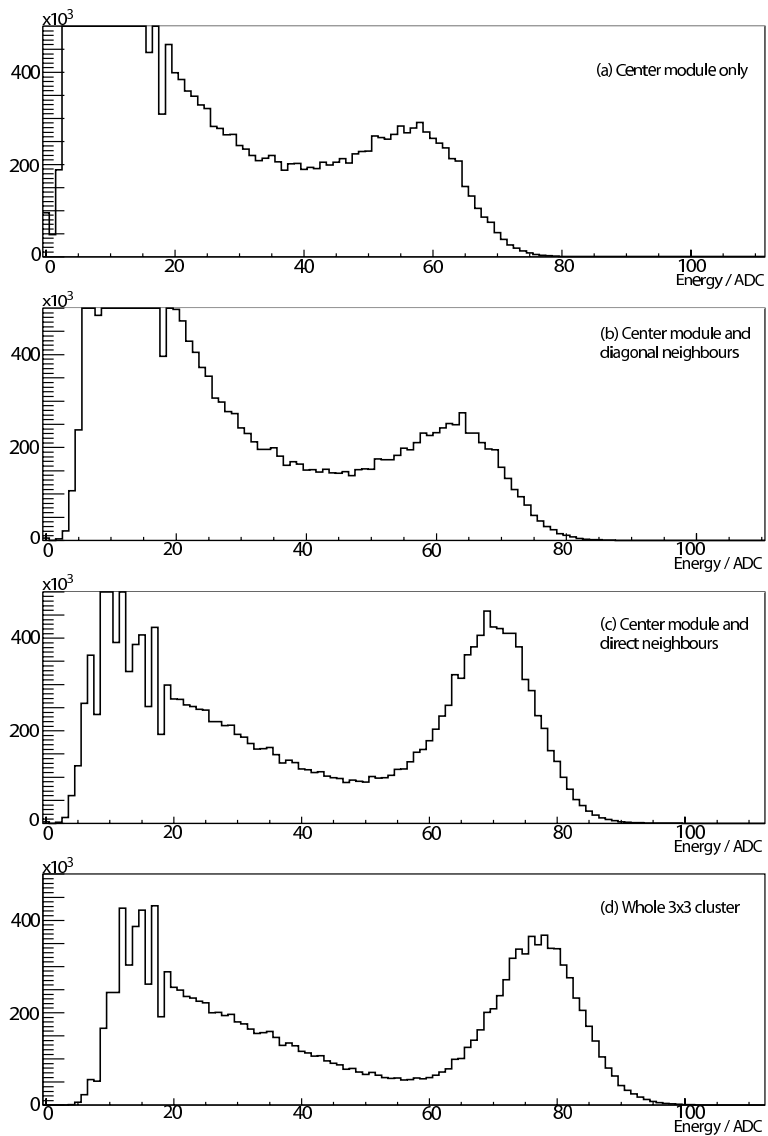}\caption{\label{fig.muebestimm}PbF$_2$
  spectra used for measuring the distribution parameters $\mu$. The beam
  energy was 510~MeV. The spectra were obtained applying the HV supply
  patterns of fig.~\ref{fig.MueMessSchema}. When the contribution of
  neighbouring modules are included, the shift of the elastic peak to higher
  ADC values is clearly visible.}
\end{center}
\end{figure}
Several of these spectra were analyzed. The offset-corrected position of the
elastic peak (see sec.~\ref{sec.SpecAna} on how the positions were determined)
is denoted by $P_{3\times3}$ when all modules were on (fig.~\ref{fig.MueMessSchema}d), $P_{cn}$ when the central
module and its direct neighbours were on (fig.~\ref{fig.MueMessSchema}c), $P_{cd}$ when central module and its
diagonal neighbours were on (fig.~\ref{fig.MueMessSchema}b )and $P_c$ when only the central module was
on (fig.~\ref{fig.MueMessSchema}a). Then one can calculate the distribution parameters as follows:
\begin{linenomath}
\begin{equation}
  \mu_c=0.95\cdot\frac{P_c}{P_{3\times3}},\,\,\,\mu_n=0.95\cdot\frac{1}{4}\frac{P_{cn}-P_{c}}{P_{3\times3}},\,\,\,\mu_d=0.95\cdot\frac{1}{4}\frac{P_{cd}-P_{c}}{P_{3\times3}}
\end{equation}
\end{linenomath}
The factor of 0.95 is used to take into account the lateral energy leakage
which is included in the calculation based on eq.~\ref{gl:modell} but which cannot
be measured by the method described here. The results are summarized in
table~\ref{tab:MueResult}. The quoted uncertainties are statistical only.
\begin{table}[htb]
  \begin{center}
       \begin{tabular}{|c|c|c|} \hline
          Parameter & Measurement & Calculation\\
          \hline
          $\mu_c$ & (64.1 $\pm$ 0.3$_{\rm{\,stat}}$) \% & 65.3 \% \\ \hline
          $\mu_n$ & ( 5.3 $\pm$ 0.1$_{\rm{\,stat}}$) \% &  6.2 \% \\ \hline
          $\mu_d$ & ( 2.3 $\pm$ 0.1$_{\rm{\,stat}}$) \% &  1.2 \% \\ \hline
       \end{tabular}   
   \end{center}
\caption{\label{tab:MueResult}Results of the measurement of the distribution
  parameters. The quoted uncertainties are statistical only. The
  calculation following~eq.~\ref{eq.DistributionPara} is based on the parameterization
  of the lateral shower distribution according to eq.~\ref{gl:modell}.}
\end{table}
The values from the measurement differ slightly but significantly from those
resulting from the calculation based on the lateral shower distribution. On
the one hand, the parameterization of the shower
distribution~eq.~\ref{gl:modell} might not describe perfectly the actual
averaged shower distribution in lead fluoride. On the other hand, systematic
errors of the measurement are not included here: the trigger conditions change
when some modules are not supplied with high voltages. Furthermore, the local
maximum (LM) and the absence of a pile-up veto is necessary for accepting an
event. These conditions are different for each of the high voltage patterns,
which adds an additional systematic uncertainty to the measurement of the
distribution parameter $\mu_c$, $\mu_n$ and $\mu_d$. Altogether, the agreement
is good enough since the differences are small enough to be irrelevant for the
calibration procedure.

\subsection{Analyzing the energy spectra}\label{sec.SpecAna}
\begin{figure}
\begin{center}
\includegraphics[scale=0.75]{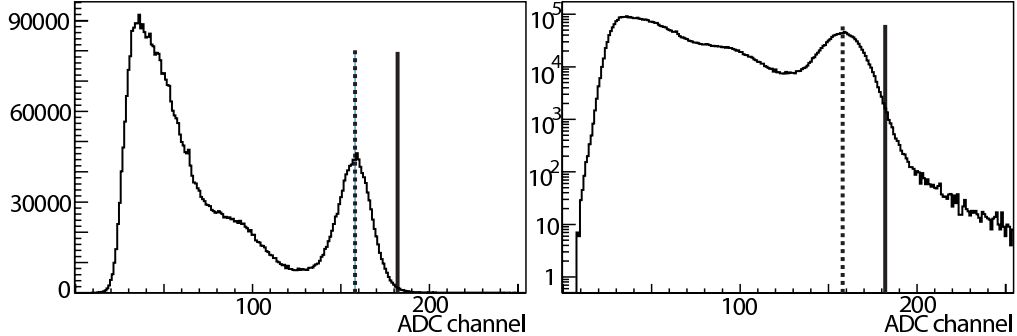}\caption{\label{fig.PeakEdge}The spots in the spectrum which we
  use to get a relationship between ADC channel and deposited energy. The
  example spectrum is plotted twice, once in linear and once in logarithmic
  scale. The position of the elastic peak is shown by the dotted line,
  the so-called high-energy ``edge'', which is defined as the inflection point at
  the right end of the logarithmic spectrum, is shown by the solid line.}
\end{center}
\end{figure}

In order to extract the information about the actual gains of calorimeter
modules, the energy spectra have to be analyzed. Although fitting is a common
method to do so, we do not use it for two reasons: first, due to the large
number of spectra per run a complete fitting to all spectra is too time
consuming for an online calibration with 10~seconds only of available time,
and second, a fit is not reliable enough because there is always the
possibility of a wrong convergence or no convergence at all. Instead, we use a
combination of filtering and peak finding algorithms. In order to establish a
relationship between ADC channels and deposited energy, we use first the
pedestals of the modules and, second, characteristic locations in the energy
spectra which are (see fig.~\ref{fig.PeakEdge}) either
\begin{itemize}
\item the elastic peak: its position in the sum spectrum
  corresponds unambiguously to a well-defined electron energy. For example, with a
  beam energy of $E=855$~MeV and a scattering angle of $\theta=35^\circ$ the
  energy of the elastically scattered electrons is $E^\prime=734$~MeV.
\end{itemize}
or
\begin{itemize}
\item the so-called ``edge'', which we define as the inflection point of the
  right shoulder in the logarithm of the spectrum. Here, the relation to a
  specific energy is only approximative.
\end{itemize}
Advantages and disadvantages of the usage of these two special locations in
the spectra will be discussed below. First we describe how these positions are
determined in the experimental spectra, which is a non-trivial task as for each five
minute run 1022 energy spectra have to be analyzed and hence a fully automatic
procedure is required.
\subsubsection{Finding the peaks in the spectra}
\label{sec.peak}
\begin{figure}
\begin{center}
\includegraphics[scale=1]{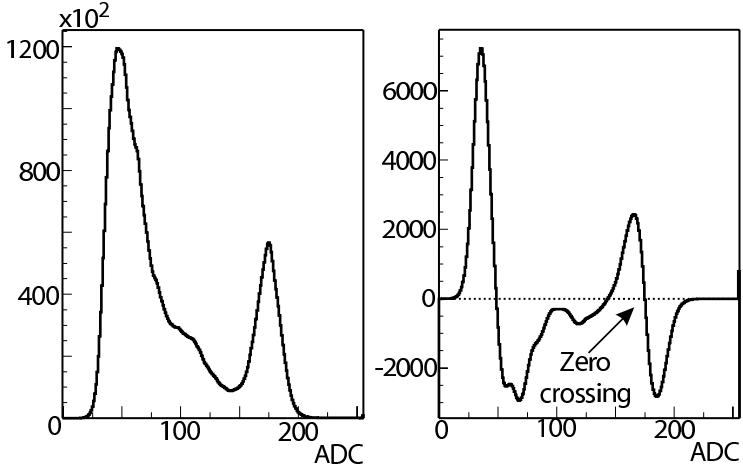}\caption{\label{fig.smoothspectrum}Energy spectrum with Gaussian
  filter applied (left) and its first derivative (right). The algorithm
  searches for the first zero crossing with negative slope starting from the
  upper end of the spectrum. This point indicates the position of the
  elastic peak in the ADC spectrum.}
\end{center}
\end{figure}
Mathematically, the elastic peak marks a local maximum in the energy spectrum
and can be identified by a zero-crossing in the first derivative. Since there are no
other maxima in the spectra at energies higher than the energy of elastically
scattered electrons, the local maximum found at the highest energy indicates
the position of the elastic peak. The procedure that determines this position
is composed of three stages and is illustrated in
fig.~\ref{fig.smoothspectrum}: first, a Gaussian filter is applied to the
spectra with a sigma of two ADC channels. This prevents misidentification of
statistical fluctuations or differential non-linearities of the ADC as a
peak. Second, the first derivative is computed by subtracting the number of
counts $N_{n}$ in the ADC channel $n$ from the number of counts $N_{n+1}$ in
the ADC channel $n+1$, i.e. $N'_n=N_{n+1}-N_{n}$, for $0\leq n \leq
254$. Third, the algorithm searches for the first zero crossing with negative
slope in the derivative starting from the highest possible ADC value, $n=254$
(and then going down to lower ADC values). Normally, for a vast majority of
calorimeter modules the positions of the elastic peaks are found
reliably. However, if the calorimeter is not well calibrated, the peaks may be
deformed to a shallow bump in the spectrum and the peak search may fail, the
treatment of these cases are discussed in section~\ref{sec.peakedge}.

\subsubsection{Finding the edges in the spectra}
\begin{figure}
\begin{center}
\includegraphics[scale=1]{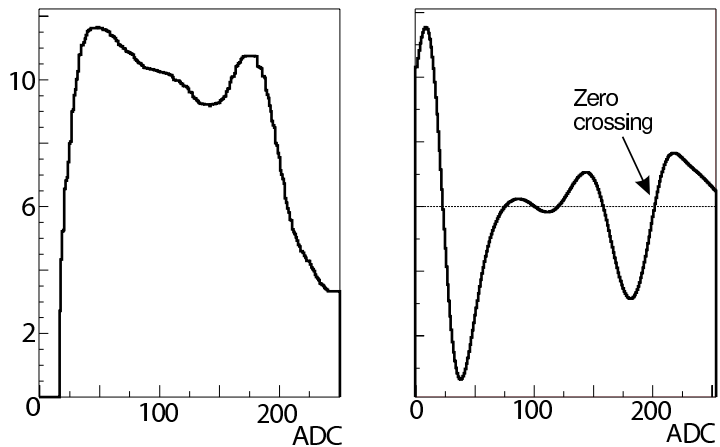}\caption{\label{fig.smoothspectrum2}Logarithm of
  an energy spectrum (left) and its second derivative (right). The indicated
  zero crossing in the second derivative marks the inflection point of the
  logarithmic spectrum which defines the position of the so-called high-energy
  ``edge''.} 
\end{center}
\end{figure}

As stated before, we define the so-called ``edge'' as the inflection point of
the right shoulder in the logarithm of the spectrum (see
fig.~\ref{fig.smoothspectrum2}). If the right side of the elastic peak was
perfectly Gaussian-shaped, there would be no inflection point in its logarithm
at all, since the logarithm of a Gaussian is a parabola.
However, the spectrum also contains a small amount of unrecognized pile-up
events with energies that may sum up to values beyond the elastic peak. This
results in an edge in the spectrum, at the location  where the number of such
pile-up events exceeds the number of elastically scattered electrons. The
procedure to find the high-energy edge is similar to that for finding the
peaks: First, the spectrum is smoothed, in this case by a median filter which
is a rank filter with a window size of 11 ADC channels in our case. The
advantage to use this kind of filter is that a median filter preserves the
position of edges whereas a Gaussian filter preserves the position of the
peaks. Next, the second derivative is computed and then the algorithm searches
for the first zero crossing with positive slope starting from the upper end of
the spectrum. Even if the calorimeter is in a completely uncalibrated state,
the high-energy edges are found with 100\% reliability.

\subsubsection{Determining the position of the elastic peak}
\label{sec.peakedge}
The position of the elastic peak together with the measured offset of the
ADC spectrum gives sufficient information to relate the sum signal of the 3$\times$3
cluster of crystals to the known energy of the elastically scattered electron. Preferably this position is
taken from the peak finding algorithm described in
section~\ref{sec.peak}. However, this peak search may fail if there are
difficult conditions like:
\begin{itemize}
\item The calorimeter is in an uncalibrated state
\item The module is located at an outer ring with only 5 neighbouring modules
\item The module is operated at its high voltage limit 
\end{itemize}
The general procedure is therefore to search both for the elastic peak and for
the high-energy edge simultaneously. If both positions are found, the
algorithm determines whether the distance between the peak and the edge is
reasonable. This distance depends on the beam energy and the energy resolution
of the cluster. If the distance exceeds specified limits, the peak
determination is assumed to have failed. In this case --- or if no peak is
found --- the peak position is estimated from the position of the high-energy
edge.

\subsection{Amplification factors of the individual modules}
With the knowledge of the distribution parameters $\mu_c$, $\mu_n$ and $\mu_d$
one can write down an equation for the signal strength $S_i$ of each
3$\times$3 cluster $i$ caused by elastic events of energy $E_0$ depending on
the amplification factors of the central crystal and its eight neighbours (see fig.~\ref{fig.cluster}):
\begin{linenomath}
\begin{equation} 
S_i=[\mu_ck_{i_0}+\mu_n(k_{i_1}+k_{i_3}+k_{i_5}+k_{i_7})+\mu_d(k_{i_2}+k_{i_4}+k_{i_6}+k_{i_8})]E_0
\end{equation}
\end{linenomath}
where $k_{i_j}$ denotes the actual amplification factor of the members $0\leq j\leq 8$ of the 3$\times$3
cluster $i$, $1\leq i \leq 1022$. The numbering for $j$ follows the definition in
fig.~\ref{fig.cluster}. Altogether there are 1022 equations for the 1022
3$\times$3 clusters of the calorimeter. Denoting from now on the
amplification factor of module $i$ by $k_i$, one can write down 1022 equations for all 1022 clusters:
\begin{linenomath}
\begin{eqnarray}
S_1&=&[\mu_ck_1+\mu_n(k_2+k_8+k_{1016})+\mu_d(k_9+k_{1017})]E_0 \nonumber\\
S_2&=&[\mu_ck_2+\mu_n(k_1+k_3+k_9+k_{1017})+\mu_d(k_8+k_{10}+k_{1016}+k_{1018})]E_0 \nonumber\\\
S_3&=&[\mu_ck_3+\mu_n(k_2+k_4+k_{10}+k_{1018})+\mu_d(k_9+k_{11}+k_{1017}+k_{1019})]E_0 \nonumber\\
\dots\nonumber\\  
S_{1022}&=&[\mu_ck_{1022}+\mu_n(k_7+k_{1015}+k_{1021})+\mu_d(k_6+k_{1014})]E_0\label{gl:signal1022}
\end{eqnarray}   
\end{linenomath}
Introducing the vectors $\vec{S}$ and $\vec{k}$,\\

\begin{linenomath}
\begin{equation}
\vec{S}=\begin{pmatrix} S_1\\S_2\\S_3\\..\\S_{1022} \end{pmatrix}
\,\,\,\,\, {\rm and}\,\,\,\,\,
\vec{k}=\begin{pmatrix} k_1\\k_2\\k_3\\..\\k_{1022} \end{pmatrix}
\end{equation}
\end{linenomath}

\noindent and the $1022 \times 1022$ matrix ${\bf A}$,
\begin{linenomath}
\begin{equation}
{\bf A}=\begin{pmatrix} \mu_c & \mu_n & 0 & 0 & 0 & 0 & 0 & \mu_n & \mu_d & ...\\
                  \mu_n & \mu_c & \mu_n & 0 & 0 & 0 & 0 & \mu_d & \mu_n & ... \\
                  0 & \mu_n & \mu_c & \mu_n & 0 & 0 & 0 & 0 & \mu_d & ...\\
                  0 & 0 & \mu_n & \mu_c & \mu_n & 0 & 0 & 0 & 0 & ... \\
                  0 & 0 & 0 & \mu_n & \mu_c & \mu_c &\mu_n & 0 & 0 & ... \\
                  \hdotsfor{10}\\
                  0 & 0 & 0 & 0 & 0 & \mu_d & \mu_n & 0 & 0 & ... 
  \end{pmatrix}
\label{gl:matrixa}
\end{equation}
\end{linenomath}

\noindent the system of linear equations can be written in matrix form
\begin{linenomath}
\begin{equation}
\vec{S}={\bf A}\vec{k}E_0
\label{gl:matrix1}
\end{equation}
\end{linenomath}


By matrix inversion of ${\bf A}$ the individual amplification factors $k_i$
of all calorimeter modules can now be calculated:

\begin{linenomath}
\begin{equation}
\vec{k}=\frac{1}{E_0}{\bf A}^{-1}\vec{S}
\label{gl:invers}
\end{equation}
\end{linenomath}
 
The inversion of a 1022$\times$1022-matrix requires usually a lot of
computation time. It is done here using a numerical method which benefits from
the fact that the matrix has non-zero elements only close to its main
diagonal. The values of the elements of the inverted matrix which are
calculated this way differ by less than one percent from those using no
approximation. The matrix ${\bf A}$ needs to be adapted to the actual
situation of the calorimeter. If, for example, the calorimeter channel $j$ is
out of order due to a technical problem, its high voltage is set to zero and
this individual channel does not contribute to the sum signal of its
neighbours. This can easily be taken into account by setting all members of
the $j$th row and $j$th column in the matrix to zero except for the diagonal
element ($j$,$j$) which is set to 1, i.e. channel $j$ is completely decoupled
from the neighbouring modules.

\subsection{Adjusting the amplification factors}
The channels are calibrated by adjusting the high voltages of the
photomultipliers. The charge $Q_i$ for a fixed number of incident photons at
the photocathode is a power function of the applied voltage $U$:
\begin{linenomath}
\begin{equation}
Q_i=\alpha_i U^{\beta_i}
\label{gl:kennlinie}
\end{equation}
\end{linenomath}
Having determined the current amplification factor $k_i$ of channel $i$ using
eq.~\ref{gl:invers}, one can calculate the new voltage $U_i^{new}$ to reach
the desired amplification factor $k_{demand}$:
\begin{linenomath}
\begin{equation}
U_i^{new}=\sqrt[\beta_i]{\frac{k_{demand}}{k_i}}U_i^{old}
\label{gl:hvbrech}
\end{equation}
\end{linenomath}

\begin{figure}
\begin{center}
\includegraphics[scale=.55]{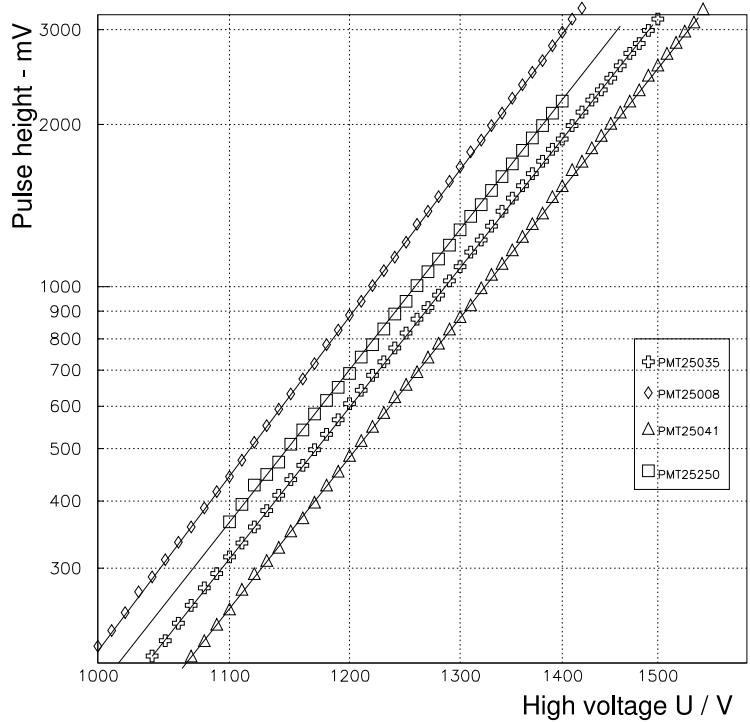}\caption{\label{fig.phototube}Pulse heights of four different
  photomultiplier tubes as a function of the applied high voltage in
  logarithmic scale. From the slope of the straight line one can
  derive the parameter $\beta$, in this case $6.48\leq\beta\leq7.39$.}
\end{center}
\end{figure}

The proportionality factor $\alpha_i$ cancels out when calculating new voltages,
so only the amplification exponent $\beta_i$ is of interest. Several tubes
have been tested using a pulsed blue LED. The measured pulse heights as a
function of the applied high voltage are shown in fig.~\ref{fig.phototube}. The
proportionality between pulse height and charge $Q$ was assumed.
Due to the large number of photomultipliers, not all $\beta_i$ have been
individually determined. Approximate values were provided by the
manufacturer. An average value of $\beta_0=7.0$ is used which means that the
error $\Delta\beta_i$ in the amplification exponent is of the order $\pm
0.5$. For $\beta_i\neq 7.0$ the new voltage $U_i^{new}$ is rather an
approximation than the exactly desired value and the calibration has to be iterated
to reach the desired charge $Q_{nom}$. It can be shown
mathematically that the procedure converges as long as the true $\beta_i\leq 1.5
\cdot \beta_0$: let $\Delta\beta$ the deviation of the the true parameter $\beta_i$ from $\beta_0$, i.e.
\begin{linenomath}
\begin{equation}
\beta_i=\beta_0+\Delta\beta
\end{equation}
\end{linenomath}
Since $Q_i$ is proportional to $k_i$, we get for the new voltage $U_i^{new}$:
\begin{linenomath}
\begin{equation}
U_i^{new}=\sqrt[\beta_i-\Delta \beta]{\frac{Q_{nom}}{Q_i^{old}}}U_i^{old}
\end{equation}
\end{linenomath}
Using eq.~\ref{gl:kennlinie}, one gets the new charge
\begin{linenomath}
\begin{equation}
Q_i^{new}=\alpha_i\left(\frac{Q_{nom}}{Q_i^{old}}\right)^{\frac{\beta_i}{\beta_i-\Delta\beta}}(U_i^{old})^{\beta_i}=\left(\frac{Q_{nom}}{Q_i^{old}}\right)^{\frac{1}{1-\frac{\Delta\beta}{\beta_i}}}Q_i^{old}
\label{gl:zwischen}
\end{equation}
\end{linenomath}
With the abbreviation
\begin{linenomath}
\begin{equation}
\gamma=\frac{\Delta \beta}{\beta_i}
\label{gl:abkuerz}
\end{equation}
\end{linenomath}
we can rewrite eq.(\ref{gl:zwischen}): 
\begin{linenomath}
\begin{equation}
Q_i^{new}=(Q_{nom})^{\frac{1}{1-\gamma}}\left(Q_i^{old}\right)^{\frac{-\gamma}{1-\gamma}}
\label{gl:zwischen2}
\end{equation}
\end{linenomath}
Division by $Q_{nom}$ yields
\begin{linenomath}
\begin{equation}
\frac{Q_i^{new}}{Q_{nom}}=\left(\frac{Q_{nom}}{Q_i^{old}}\right)^{\frac{\gamma}{1-\gamma}}
\label{gl:warum}
\end{equation}
\end{linenomath}
From eq.~\ref{gl:warum} follows that the calculation of new voltages converges
as long as
\begin{linenomath}
\begin{equation}
|\frac{\gamma}{1-\gamma}|<1
\label{gl:soso}
\end{equation}
\end{linenomath}
Hence, the convergence constraint for $\gamma$ and $\Delta\beta$ respectively is:
\begin{linenomath}
\begin{equation}
\gamma<\frac{1}{2} \quad\Leftrightarrow\quad \Delta\beta<\frac{1}{2}\beta_i 
\end{equation}
\end{linenomath}
The condition $\Delta\beta<\frac{1}{2}\beta_i$ means in our case that as long
as $\Delta\beta<3.5$, the procedure still converges. This requirement is very
well fulfilled since for our photomultipliers
$\,\,-\frac{1}{2}<\Delta\beta<+\frac{1}{2}$, so one can expect a fast convergence.

\section{Measurements and results}\label{sec.Measurements}
Since the setup of the PbF$_2$ calorimeter in 2000, the calibration procedure has
been successfully used during 8000 hours of beam. In this section, some of the
benefits and properties will be presented.

\subsection{Calibration starting from an uncalibrated state}
\begin{figure}
\begin{center}
\includegraphics[scale=1.3]{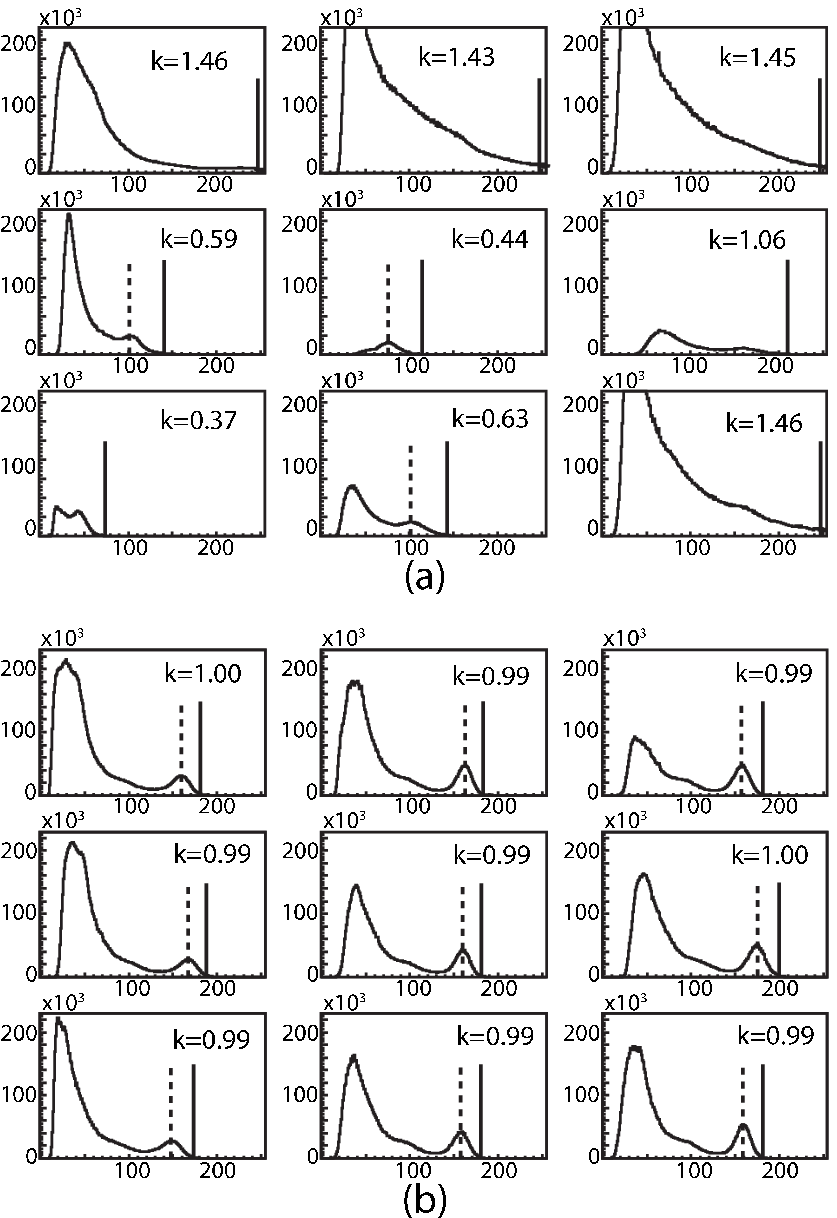}\caption{\label{fig.3x3cluster}Energy
  spectra of 3$\times$3 crystal clusters in uncalibrated state~(a) and
  after calibration~(b). The positions of the high energy edge that were found
  by the algorithm are indicated by the solid lines. The positions of the
  elastic peak --- if found by the algorithm --- are indicated by the dashed
  lines. The normalized amplification factors $k=k_i/k_{nom}$ for each module
  are denoted in the graphs. After three calibration steps the spectra get
  more uniform and the elastic peaks become more prominent.}
\end{center}
\end{figure}

To test the calibration procedure, the calorimeter was brought in an
un\-calibrated state by applying random high voltages within the allowed
limits to all modules. An amplification factor $k_{nom}$ was demanded so that
the elastic peak lies at ADC channel 170 above pedestals. In this
case, three steps were sufficient to bring the calorimeter into a
well-calibrated state. Fig.~\ref{fig.3x3cluster} shows 3$\times$3 sum spectra
before and after calibration. One can clearly see the enhancement in energy
resolution and uniformity of the energy spectra. 
\begin{figure}
\includegraphics[scale=0.55]{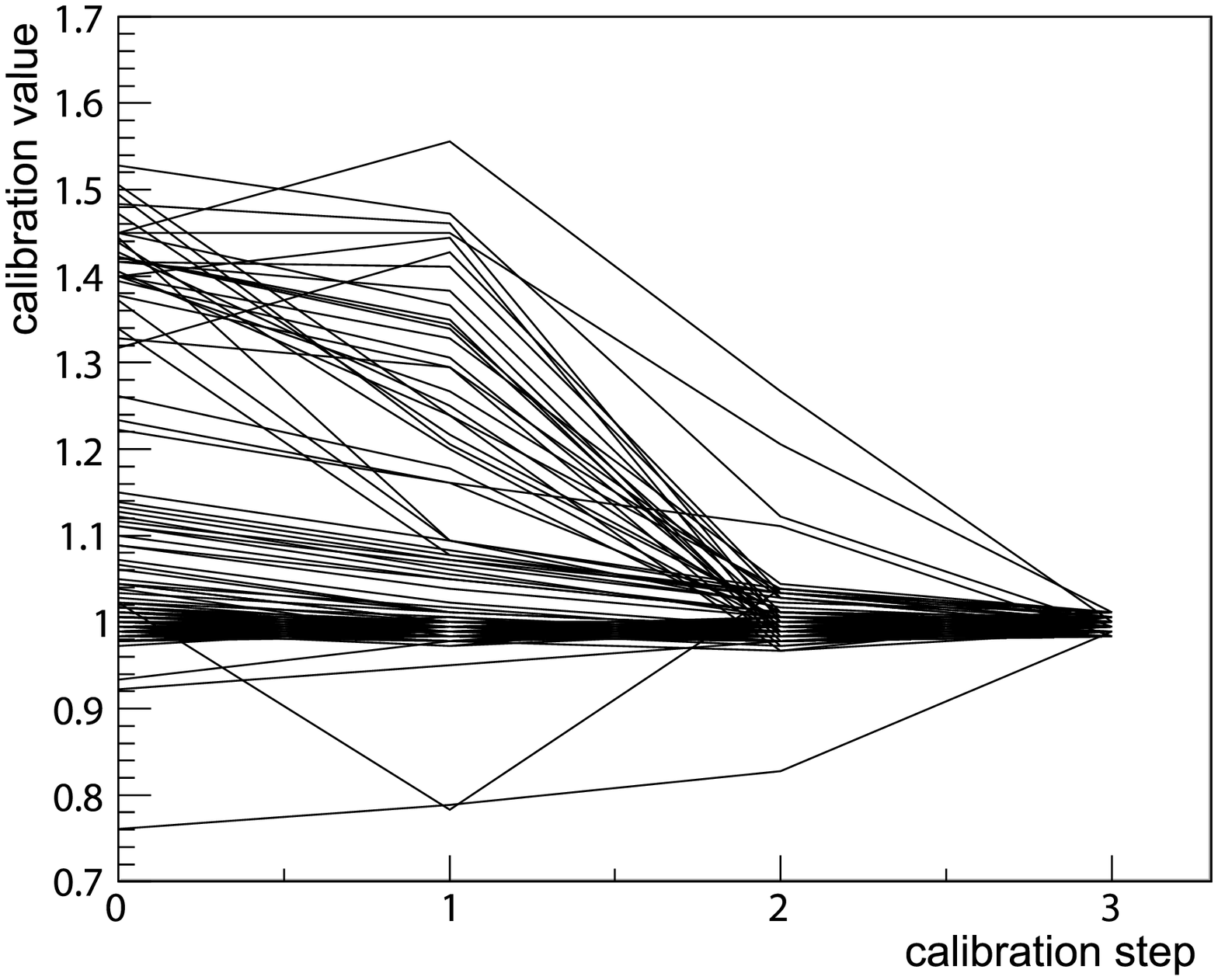}\caption{\label{fig.calibconst}Normalized
  amplification factors $k$ of selected single modules and their development during
  the calibration procedure. After three calibration steps the mean
  calibration value changed from 1.02 to 1.00, the standard deviation decreased
  from 0.11 to 0.01.}
\end{figure}
Fig.~\ref{fig.calibconst} shows how the normalized amplification factors
$k=k_i/k_{nom}$ of the single modules evolve during the calibration
procedure. A normalized amplification factor of $k_i=1$ means that channel $i$
is perfectly calibrated. After three calibration steps the mean calibration
value changed from $\bar{k}=1.02$ to $\bar{k}=1.00$, the standard deviation decreased
from 0.11 to 0.01.

\subsection{Energy resolution of the PbF$_2$ calorimeter}
To quantify the energy resolution of the calorimeter, we use the width of the
elastic peaks in the measured energy spectra. The observed peak width depends
on various factors, like the variation of the scattered electron energy within
the acceptance, energy straggling of the electrons within the hydrogen target
and the material between target and calorimeter crystals, or the intrinsic
energy resolution of the detector material itself.\\

This observed relative energy resolution is determined by fitting a phenomenological
function to the elastic peak of the energy spectra~\cite{Matulewicz90} which
takes into account the radiative tail and the shower leakages. The function
describes the peak with a Gaussian with mean $\mu$ and width $\sigma_R$ for
the high energy side and a Gaussian with mean $\mu$ and width $\sigma_L$
modified by an exponential for the low energy side:

\begin{linenomath}
\begin{equation}
  f(x)= \left\{ 
  \begin{array}{l l r}
    C\cdot \left\{\exp\left[-\frac{1}{2}\left(\frac{x-\mu}{\sigma_L}
      \right)^2\right]+\exp\left[\frac{x-\mu}{\lambda}\right]\cdot\left
    (1-\exp\left[-\frac{1}{2}\left(\frac{x-\mu}{\sigma_L}\right)^2\right]\right)\right\}&\mathrm{,}&x<\mu\\
    C\cdot
    \exp\left[-\frac{1}{2}\left(\frac{x-\mu}{\sigma_R}\right)^2\right]&\mathrm{,}&x\geq\mu
  \end{array}
  \right. 
  \label{gl:FitPeak}
\end{equation}
\end{linenomath}
The relative energy resolution $\Delta E/E$ is defined here as $\sigma_R/\mu$.
Because the signals of nine neighbouring channels are summed up, the
calibration enhances the energy resolution. For the calibration example
presented in the previous section, the average effective resolution improved
from $13.9\%$ before calibration to $5.0\%$ after calibration.

\subsection{Stability of the calibration and routine operation}
\begin{figure}
\begin{center}
\includegraphics[scale=.8]{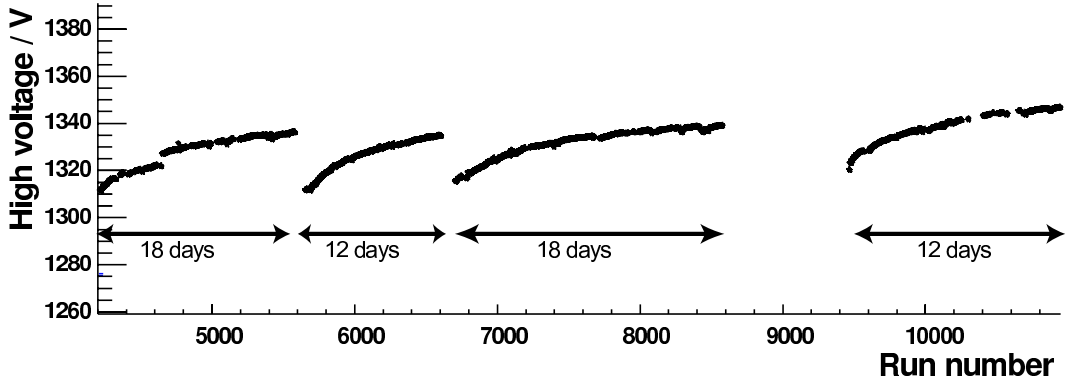}\caption{\label{fig.HVvsRunno}Average
  PMT high voltage of the calorimeter versus run number. The graph covers four beam
  times where one can see an increase of the voltage with time because of the
  radiation damage in the PbF$_2$ crystals. After each beam time the radiation
  damage is reduced by exposing the crystals to blue light thus reducing the
  necessary HV again.}
\end{center}
\end{figure}

Once a calibrated state has been reached, the calorimeter remains calibrated
for several hours. In order to test the stability, we left the high
voltage of the multipliers unchanged during a measurement period of 24~h and observed
only small decalibration effects. This is possible because of the stable
experimental conditions: the temperature in the experimental halls is kept
stable to $\Delta T=\pm 1^{\circ}$C, the beam current is stabilized on the
$10^{-6}$ level and hence the rates on the detector are very stable. The main
source of decalibration is the response of the PbF$_2$ crystals to the high
radiation level to which they are exposed. Radiation damages in the crystals
lead to the formation of color centers and thus to a loss of light yield which is
specific to each crystal. The calibration procedure compensates for this
effect by increasing the high voltages of the photomultiplier tubes and is
applied once per hour. During a beam period of three weeks, the average high
voltage is increased by about 25~V.  After a data taking period (typically two
weeks), optical bleaching is applied to the crystals~\cite{Achenbach98} and
most of the radiation damages are repaired. The average HV at the beginning
of the following beam time is then lowered again, as shown in fig.~\ref{fig.HVvsRunno},
where the mean voltage of the calorimeter channels is shown as a function of
the run number. Ten runs correspond to one hour of data taking. One can see
four periods of data taking, each of them 2--3~weeks long. The high voltages
start at low values and are increased automatically by the calibration
procedure to compensate for the radiation damage.\\ Using massive
parallelization and high-speed network equipment, the interval between runs
needed for readout, storage and preparation of the next run was reduced to
30~s, resulting in a data taking efficiency of 90\%. The calibration procedure
described here can be partly performed during data taking, applying the new
high voltages can be done in the time window between two runs without
compromising the efficiency.

\section{Summary}
A calibration procedure was developed for a segmented, total absorbing
homogenous electromagnetic calorimeter consisting of 1022~lead fluoride
crystals. Since the signals of 3$\times$3 neighbouring modules are summed up
and digitized to determine the energy of a single event, a uniform gain in all
modules is a prerequisite for a good effective energy resolution. In order to
achieve this, the photomultipliers were precalibrated before they were
installed in the calorimeter. For the analog summation circuit high precision
components such as resistors with 0.1\% tolerance and capacitors with 1\%
tolerance were used. A uniform gain is then obtained by the calibration
procedure that was presented here. Three conditions had to be fulfilled:
first, the large number of channels required a fully automatic, reliable
treatment. Second, since only the sum signals are available with sufficient
digital resolution, the amplification factors of the single modules had to be
extracted from the sum signal. And third, the method had to tolerate ignorance
of the precise gain coefficients of the PMTs.

The first issue was addressed by implementing a combined search for the
elastic peak and the so-called high-energy ``edge'' in the energy spectra which
ensures that the calibration state of the 3$\times$3-clusters is always correctly
determined. For the second issue a mathematical solution was found by forming
a set of 1022 linear equations and solving them by inverting a 1022$\times$1022
matrix. For the third issue it could be proven that the exact knowledge of the
individual exponential gain factors $\beta$ of the photomultipliers is not necessary as long as the true
gain factor of a multiplier does not exceed the average gain factor which is used for the
calculation of the voltages by more than a factor of 1.5.\\

The calibration procedure has been in use for several years now. More than
$10^{8}$ energy spectra measured at beam energies from 300~MeV up to 1500~MeV
have been successfully analyzed. The calorimeter can be brought from an
uncalibrated to a well-calibrated state within three calibration steps
corresponding to 10~minutes real time. The energy resolution is significantly
enhanced within these three calibration steps and can be kept at the same
level by applying a new calibration step once an hour.

\section{Acknowledgements}
This work was supported by the Deutsche Forschungsgemeinschaft in the
framework of the CRC 201, 443 and the SPP 1034. We would like to thank the
crew of the MAMI accelerator for providing us with an excellent electron beam.

\end{document}